\documentclass[aps,pre,reprint,superscriptaddress]{revtex4-1}

\usepackage{graphicx}
\usepackage{bm}
\usepackage{color}
\usepackage{amsmath}
\usepackage{amssymb}
\usepackage{mathtools}
\usepackage{dsfont}
\usepackage{color}
\usepackage{calc}
\usepackage{hyperref}
\usepackage{multirow}
\usepackage{natmove}
\usepackage{epstopdf}
\newcommand{\lla}{\left\langle}
\newcommand{\rra}{\right\rangle}

\graphicspath{{./figures/}}

\begin{document}
\title{Bacterial swarmer cells in confinement: A mesoscale hydrodynamic simulation study}

\author{Thomas Eisenstecken}
\email{t.eisenstecken@fz-juelich.de}
\affiliation{Theoretical Soft Matter and Biophysics, Institute for
Advanced Simulation and Institute of Complex Systems,
Forschungszentrum J\"ulich, D-52425 J\"ulich, Germany}
\author{Jinglei Hu}
\email{hujinglei@nju.edu.cn}
\affiliation{Kuang Yaming Honors School, Nanjing University, 210023 Nanjing, China}
\author{Roland G. Winkler}
\email{r.winkler@fz-juelich.de}
\affiliation{Theoretical Soft Matter and Biophysics, Institute for
Advanced Simulation and Institute of Complex Systems,
Forschungszentrum J\"ulich, D-52425 J\"ulich, Germany}


\date{\today}

\begin{abstract}
A wide spectrum of {\em Peritrichous} bacteria undergo considerable physiological changes when they are inoculated onto nutrition-rich surfaces and exhibit a rapid and collective migration denoted as swarming. Thereby, the length of such swarmer cells and their number of flagella increases substantially. In this article, we investigated the properties of individual {\em E. coli}-type swarmer cells confined between two parallel walls via mesoscale hydrodynamic simulations, combining molecular dynamics simulations of the swarmer cell with the multiparticle particle collision dynamics approach for the embedding fluid.
{\em E. coli}-type swarmer cells are three-times longer than their planktonic counter parts, but their flagella density is comparable. By varying the wall separation, we analyze the confinement effect  on the flagella arrangement, on the distribution of cells in the gap between the walls, and on the cell dynamics. We find only a weak dependence of confinement on the bundle structure and dynamics. The distribution of cells in the gap changes from a geometry-dominated behavior for very narrow to fluid-dominated behavior for wider gaps, where cells are preferentially located in the gap center for narrower gaps and stay preferentially next to one of the walls for wider gaps. Dynamically, the cells exhibit a wide spectrum of migration behaviors, depending on their flagella bundle arrangement, and ranges from straight swimming to wall rolling. 
\end{abstract}
\pacs{}
\keywords{}
\maketitle

\section{Introduction}

Many motile bacteria are propelled by helical filaments, which protrude from their cell body and are driven by rotary motors located in the cell membrane.  \cite{berg:72,berg:73,berg:04,tayl:51} Thereby, such bacteria exhibit different modes of locomotion, depending on the environment. In liquid environments, individual (planktonic) cells exhibit the so-called {\em swimming} motility.\cite{henr:72,kear:10,darn:10,turn:10} The various flagella
of {\em peritrichous} bacteria self-organize into bundles by
(typically) counterclockwise rotation of the flagella motors. This leads to  nearly straight swimming in bulk fluids and circular motion near walls.\cite{laug:06,dile:11,leme:13,hu:15} To change the swimming direction,
this ``running'' phase is interrupted by short periods of
``tumbling''.\cite{turn:00,darn:07.1,wata:10,berg:72,plat:97,scha:02,schm:02,xie:11,reig:12,reig:13,adhy:15} The sequence of run-and-tumble events can be adjusted by chemotaxis, i.e.,  in response to changes in chemical concentrations.\cite{darn:10}

Another mode of motion is denoted as bacterial {\em swarming}, where flagellated bacteria migrate collectively over surfaces and are able to form stable aggregates, which can become highly motile.\cite{henr:72,kear:10,darn:10,turn:10,arie:15} Swarming bacteria show a strikingly different motile behavior than swimming cells. They are densely packed and exhibit large-scale swirling and streaming motions. Some  bacteria strains show distinctly different morphologies in the swarming mode compared to the swimmer cells as they are more elongated by suppression of cell division and their number of flagella is significantly increased. \cite{stah:83,jone:04,cope:09,kear:10,darn:10,tuso:13} This
points toward the significance of flagella for swarming. {\em E. coli} and {\em Salmonella} bacteria more than double their length and increase the number of flagella, but the flagellar density remains approximately constant.\cite{cope:10,kear:10,part:13,part:13.1,swie:13} The changes for {\em P. mirabilis} are even more dramatic, their length increases $10$ to $50$ times and an increase of their flagella number from fewer than $10$ to $5 000$ has been  reported. \cite{cope:09,mcca:10,tuso:13} As stated in Ref.~\onlinecite{kear:10}, neither is the reason known why swarming requires multiple flagella nor why a significant cell elongation is required for many bacteria. Aside from a possible amplification of swarming by shape-induced alignment of adjacent cells, elongation
associated with the increase in the number of flagella may help to overcome wall friction. \cite{part:13}

\begin{figure*}[t!]
\centering
  \includegraphics[height=\textwidth,angle=90]{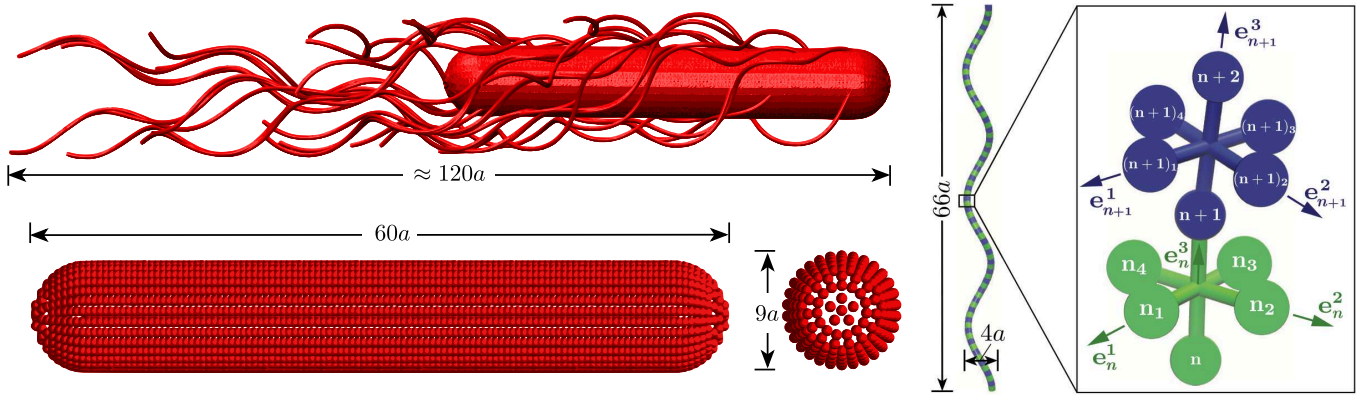}
  \caption{Model of the spherocylindrical cell body and the three-turn left-handed flagella. The cell body is three times longer than that of swimmer cell. The anchoring points of the 25 flagella on the body surface are randomly chosen. The flagellum, a three-turn left-handed helix, consists of 76 consecutive segments.  In each segment, six particles are arranged in an octahedron.\cite{hu:15.1}}
  \label{fig:cell}
\end{figure*}

This brief list already indicates that very little is known about the locomotion of  swarming bacteria and their interactions. To the best of our knowledge, no theoretical study of individual swarmer cells has be performed so far, even less their collective behavior. The reason is twofold. On the hand, an adequate model of a multi-flagellated bacteria is required. On the other hand, hydrodynamic interactions have to be taken into account. The complexity of the propulsion mechanism with bundle formation of flagella, especially near-field hydrodynamics, poses substantial challenges for simulations.\cite{reig:12,reig:13,hu:15} Both aspects are demanding in terms of computational resources and require the simulation of large systems.

In this article, we adopt a mesoscale hydrodynamic simulation approach to study the properties of individual swarmer cell in thin films as a first step to unravel their specific motility properties during swarming. We extent our previous bacterium model,\cite{hu:15.1} which closely resembles the geometry, flagellar elastic properties, and rotary motor torque of {\em E. coli} to a multi-flagellated swarmer cell. The fluid is modelled by the multiparticle collision dynamics (MPC) method, a particle-based simulation approach taking into account hydrodynamic interactions and thermal fluctuations.\cite{male:99,kapr:08,gomp:09} The MPC method has proven to be very valuable for the studies of active systems.\cite{ruec:07,goet:10,yang:11,kapr:08,elge:09,earl:07,elge:10,babu:12,elge:13,thee:13,yang:14,zoet:14,elge:15,hu:15,hu:15.1} Specifically, MPC has successfully been applied to elucidate synchronization between the flagella beating of nearby swimming sperm,\cite{yang:08} bundling of helical flagella of bacteria,\cite{reig:12,reig:13} and swimming of bacteria near walls.\cite{hu:15}

Our {\em E. coli}-type swarmer cell is three times longer than a planktonic cell and is covered with 25 flagella. We consider various realizations with randomly arranged flagella, and find, in general, rather heterogeneous properties. One of our goals is to shed light on the organization of the flagella into bundles. We find that the majority of flagella self-organize into a major bundle, essentially independent of the extend of confinement. The bundle and the cell body exhibit a pronounced angle for most surfaces separations. Such a structure has already be seen experimentally for a planktonic {\em E. coli} cell.\cite{darn:07.1} In very narrow slits, the distribution of cells strongly depends on their bundle arrangement, and cells may preferential be very close to the walls. For slightly wider gaps, the configuration with cells in the center between the walls is preferred, and for very wide gaps, cells migrate preferentially along one wall. Most cells move essentially in a straight manner in narrow gaps. Some cells exhibit a  more complicated dynamics in wide gaps and  roll over a wall.

An important conclusion of our studies is that the considered type of swarmer cell is rather similar to a swimmer cell as far as the flagella bundle characteristics and migration behavior of individual cells is concerned. Of course, such {\em E. coli}-type swarmer cells exhibit swarming at surfaces. However, they lack specificities of long swarmer cells such as multiple bundles as indicated in the images of Refs.~\onlinecite{stah:83} and \onlinecite{jone:04}. We expect that such multiple bundles will give rise to additional collective effects, with qualitative and quantitative differences to the (short) considered swarmer cell.

\section{Model of Swarmer and Fluid}

\subsection{Swarmer Model}

We use an extension of the bacteria model described in Refs.~\onlinecite{hu:15} and \onlinecite{hu:15.1}. The swarmer cell is composed of a spherocylindrical body and attached flagellar filaments, as illustrated in Fig.~\ref{fig:cell}. Both, the body and the flagella are constructed by connected mass points of mass $M$. The spherocylinder consists of circular sections, each with a center particle and  uniformly distributed particles on its circumference.  The larger circles comprise $30$ particles, whereas the smaller ones toward the poles consist of 15 and 5 particles, respectively. In order to maintain the shape of the body, nearest- and next-nearest-neighboring pairs of particles are connected by a harmonic potential of the form
\begin{equation}  \label{Eq:Vbd}
U_{\rm b}=\dfrac{1}{2}\,K_{\rm b}(r-r_{\rm e})^2 ,
\end{equation}
where $r$ and $r_e$ are the distance between the respective pair and its preferred (equilibrium) value. Moreover, the circle-center particle is connected similarly with every particle at the circumference as well as its neighboring center particles.

A flagellum is described by the helical wormlike chain model, \cite{yama:97,voge:10,brac:14} with an adaptation suitable for the combination with MPC.\cite{hu:15.1} As shown in Fig.~\ref{fig:cell}, a helical flagellum consists of $N_F = 76$ octahedron-like segments with a total of $381$ particles. In each segment, six particles are arranged in an octahedron of edge length $a/\sqrt{2}$, forming $12$ bonds along the edges and three along the diagonals, where $a$ is the unit length of the MPC fluid as described in Sec.~\ref{sec:mpc}.. The preferred bond lengths are $r_{\rm e} = a/\sqrt{2}$ for edges and $r_{\rm e} = a$ for diagonals. This construction allows for a straightforward description of the intrinsic twist of a flagellum and a coupling of the twist to the forces exerted by the MPC fluid.

The bond vectors ${\bm b}_n^3 = {\bm r}_{n+1} - {\bm r}_{n}$ ($n = 1,...,N_F$) specify the backbone of the flagellum, and, together with ${\bm b}_{n}^1 = {\bm r}_{n_1} - {\bm r}_{n_3}$ and ${\bm b}_n^2 = {\bm r}_{n_2} - {\bm r}_{n_4}$, define orthonormal triads $\{ {\bm e}_n^1,\,{\bm e}_n^2,\,{\bm e}_n^3\}$, where ${\bm e}_n^\alpha = {\bm b}_n^\alpha / \vert{\bm b}_n^\alpha \vert$,  $\alpha \in \{1, 2, 3\}$. Here,  ${\bm r}_{n}$  denotes the position of the backbone particle $n$, and the ${\bm r}_{n_k}$ ($k =1,\ 2,\ 3,\ 4$) refer to the positions of the particles in the plane with the normal  ${\bm e}_n^3$ (cf. Fig.~\ref{fig:cell}).

The local elastic deformation of a flagellum is characterized by the transport of  the triad $\{ {\bm e}_n^1,\, {\bm e}_n^2,\, {\bm e}_n^3 \}$ to $\{ {\bm e}_{n+1}^1,\, {\bm e}_{n+1}^2,\, {\bm e}_{n+1}^3 \}$ along the helix.\cite{voge:10} This process is performed in two steps: (i) the rotation of $\{ {\bm e}_n^1,\, {\bm e}_n^2,\, {\bm e}_n^3 \}$ around ${\bm e}_n^3$ by a twist angle $\varphi_n$, and (ii) the rotation of the twisted triad $\{ {\tilde{\bm e}}_n^1,\, {\tilde{\bm e}}_n^2,\, {\tilde{\bm e}}_n^3 \}$ by a  bending angle $\vartheta_n$ around the unit vector ${\bm n}_n = ({\bm e}_n^3 \times {\bm e}_{n+1}^3) / \vert {\bm e}_n^3 \times {\bm e}_{n+1}^3 \vert$ normal to the plane defined by the contour bonds ${\bm b}_n^3$ and ${\bm b}_{n+1}^3$. The corresponding elastic deformation energy is
\begin{equation}
U_{\rm el} = \dfrac{1}{2}\sum_{\alpha=1}^3K_{\rm el}^\alpha\sum_{n=1}^{N-1}(\Omega_n^\alpha - \Omega_{\rm e}^\alpha)^2,
\label{Eq:Vel}
\end{equation}
where $K_{\rm el}^1=K_{\rm el}^2$ is the bending strength, $K_{\rm el}^3$ the twist strength, and  ${\bm{\Omega}}_n= \sum_{\alpha} \Omega_n^\alpha \bm{ e}_n^\alpha  = \vartheta_n {\bm n}_n + \varphi_n {\bm e}_n^3$ the strain vector. The parameters $\Omega_{\rm e}^\alpha$  define the equilibrium geometry of the model flagellum and are chosen to recover the shape of an {\em E. coli} flagellum in the normal state, i.e., a three-turn left-handed helix. \cite{darn:07.1}

\begin{figure}[t]
\centering
  \includegraphics[width=0.45\textwidth]{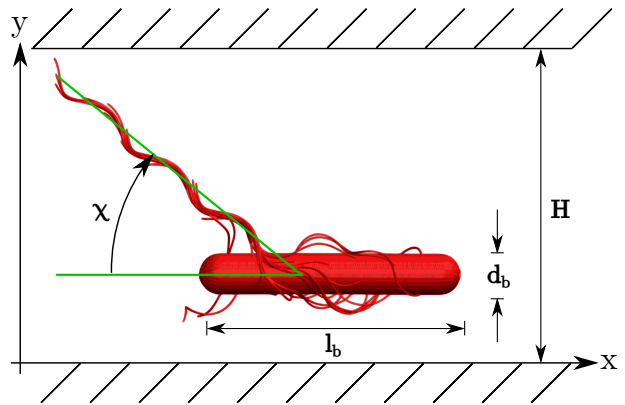}
  \caption{Illustration of the simulation set-up. A swarmer cell is confined between two walls parallel to the $xz$-plane separated by $H$. }
  \label{fig:sim_angle}
\end{figure}

We do not explicitly model the hook connecting a flagellum with the cell body of a bacterium,\cite{wata:10} but rather directly attach a flagellum to the cell body by choosing a body particle as its first contour particle ($n=1$, see Fig.~\ref{fig:cell} for notation). To induce rotation of the flagellum, a motor torque ${\bm T}$ is applied, which is decomposed into a force couple ${\bm F}$ and $-{\bm F}$ acting on particles $1_2$ and $1_4$ (${\bm T}={\bm b}_1^2\times {\bm F}$ with ${\bm F}$ antiparallel to ${\bm b}_1^1$), or equivalently $1_1$ and $1_3$ (${\bm T}={\bm b}_1^1\times {\bm F}$ with ${\bm F}$ parallel to ${\bm b}_1^2$). Hence, the bacterium is force free. To ensure that the bacterium is also torque-free, an opposite torque $-{\bm T}$ is applied to the body. Penetration of a flagellum into the cell body and crossing of flagella is prevented by the harmonic repulsive potential
\begin{equation} \label{eq:rep_pot}
U_{\rm ex} = \left\{\begin{array}{cc}
\dfrac{1}{2}\,K_{\rm ex}(r-r_{\rm ex})^2 , & r< r_{\rm ex} \\
0 , & \text{otherwise}
\end{array} .
\right.
\end{equation}
For the flagellum-body interaction, we consider the repulsion with the body-center particles only in order to reduce the numerical effort. Hence, we set for these interactions $r_{\rm ex} = (d_b+a)/2$, where $d_b$ is the diameter of the cell body (cf. Figs.~\ref{fig:cell} and \ref{fig:sim_angle}).  In case of the flagellum-flagellum repulsive interaction, $r$ is the closest distance between contour bonds of flagella and $r_{\rm ex}= 0.25\, a$.\cite{kuma:01}

The dynamics of the bacterium is described by Newton's equations of motion with the forces resulting from the potentials of Eqs. (\ref{Eq:Vbd})--(\ref{eq:rep_pot}) and the ``external'' forces for generating the torques ${\bm T}$ and $-{\bm T}$.

\subsection{Fluid: Multiparticle Collision Dynamics} \label{sec:mpc}

In MPC, the fluid is described by $N$ point particles of mass $m$ with continuous positions
${\bm r}_i$ and velocities ${\bm v}_i$ ($i=1,\ldots, N$), which interact with each other by a stochastic, momentum-conserving process. The particle dynamics proceeds in a sequence of streaming and collision steps. In the ballistic streaming step, the particle positions are
updated according to
\begin{align} \label{streaming}
 {\bm r}_i(t+h)={\bm r}_i(t)+ h {\bm v}_i(t),
\end{align}
where $h$ is the collision time step. In the collision step, the MPC particles are sorted into cubic collision cells of length $a$, which define the local multiparticle collision environment. In the stochastic rotation dynamics (SRD) version of MPC, \cite{male:99,kapr:08,gomp:09} the relative velocity of each particle, with respect to the center-of-mass velocity of the collision
cell, is rotated by a fixed angle $\alpha$ around a randomly oriented axis.
Hence, the velocities after the stochastic interaction are given by \cite{thee:16}
\begin{align} \label{eq:coll_angular}
 {\bm v}_i(t+h)  & =   {\bm v}_{cm}(t)+\mathbf{R}(\alpha) [{\bm v}_i(t)- {\bm v}_{cm}(t)] - {\bm r}_{i,c} \times \\ \nonumber
	  & \Big[ m \mathbf{I}^{-1} \sum_{j \in cell}\left\{{\bm r}_{j,c}(t)\times \left[{\bm v}_{j,c}(t)-\mathbf{R}(\alpha)  {\bm v}_{j,c}(t) \right] \right\} \Big],
\end{align}
where
\begin{align} \label{center-of-mass}
{\bm v}_{cm}=\frac{1}{N_c}\sum_{i=1}^{N_c} {\bm v}_i
\end{align}
is the center-of-mass velocity, $N_c$ the total number of particles in the collision cell,
$\mathbf{I}$ the moment-of-inertia tensor of the particles in the center-of-mass reference frame, and
${\bm r}_{i,c} (t)$ and ${\bm v}_{i,c} (t)$ are the  relative positions and velocities after streaming, i.e.,
${\bm r}_{i,c}  ={\bm r}_{i}-{\bm r}_{cm}$ and ${\bm v}_{i,c}={\bm v}_{i}-{\bm v}_{cm}$, with the center-of-mass position ${\bm r}_{cm}$. The collision rule (\ref{eq:coll_angular}) conserves angular momentum on the collision cell level by  a solid-body type  rotation of relative velocities after a  collision.\cite{nogu:08,thee:14,thee:16}
In its original version, MPC violates Galilean
invariance. It is restored by a random
shift of the collision grid at every step.\cite{ihle:03}
In order to simulate an isothermal fluid, a collision-cell-based, local Maxwellian thermostat is applied,
where the relative velocities of the particles in a collision cell are scaled according to the
Maxwell-Boltzmann scaling (MBS) method.\cite{huan:10,huan:15}

Since the MPC algorithm is highly parallel,  we exploit a graphics processor unit (GPU)-based version of the simulation code, which yields a high performance gain.\cite{west:14}

\subsection{Coupling of Bacterium and MPC Fluid}

The coupling between the MPC particles and the mass points of the bacterium body and flagella is efficiently achieved in the MPC collision step.\cite{male:00.1,ripo:04,muss:05} Thereby, the cell points are treated on equal footing with the MPC particles, i.e., their velocities are also rotated according Eq.~(\ref{eq:coll_angular}) to ensure momentum exchange between them and the fluid. The center-of-mass velocity of a collision cell containing mass points of a cell is now given by
\begin{align} \label{center-of-mass_cell}
{\bm v}_{cm}=\frac{1}{mN_c+ M N_c^c} \left( \sum_{i =1}^{N_c} m {\bm v}_i + \sum_{j=1}^{N_c^c} M {\bm v}_j^b \right) .
\end{align}
Here, $N_c^c$ is the number of mass points of a bacterium in the considered collision cell.

\subsection{Wall Interactions}

Our swarmer cells are confined between two walls, which are parallel to the $xz$-plane and separated by a distance $H$ (cf. Fig.~\ref{fig:sim_angle}). Various wall separations are considered, ranging from $H/a=20$ to $H/a=120$, or in units of body length from $H/l_b =1/3$ to $H/l_b=2$. No-slip boundary conditions are applied for the MPC fluid at the walls by implementing the bounce-back rule and taking into account phantom particles in the walls. \cite{lamu:01,huan:15} The mass points of a cell experience the reflecting Lennard-Jones potential (wall at $y=0$)
\begin{align}
U_{w} = \left\{
\begin{array}{cc}
4 k_BT \left[ \left( \displaystyle \frac{\sigma}{y-R} \right)^{12} - \displaystyle \left( \frac{\sigma}{y-R} \right)^6  \right]  , & y -R < y_c \\
0 , & \text{otherwise}
\end{array} .
\right.
\end{align}
Here, $y$ is either the distance between a flagellum contour particle and the wall, or that of a body-center particle and the wall. Hence, we set $R=0$ for the flagella particles and $R=d_b/2$ for the cell body. The cut-off distance is $y_c=\sqrt[6]{2}a$.

Initially, cells with randomly oriented and randomly anchored flagella are placed in the narrowest channel and are partially equilibrated until a loose bundle is formed as exemplified in Fig.~\ref{fig:cell}; in total, we consider $11$ distinct realizations.
These structures are utilized for the studies of all gap widths, where the swarmer cells are further equilibrated with different starting velocities of the swarmer and fluid particles. Thereby, only one of the walls is displaced, i.e., the cells are initially close to one of the walls.  In general, we find a significant heterogeneity in the appearing structures and the dynamical properties of individual cells. This is consistent with experimental observation of the properties of {\em E. coli} bacteria.\cite{chat:06,darn:07.1}

\begin{figure}[t]
\centering
  \includegraphics[width=0.95\columnwidth]{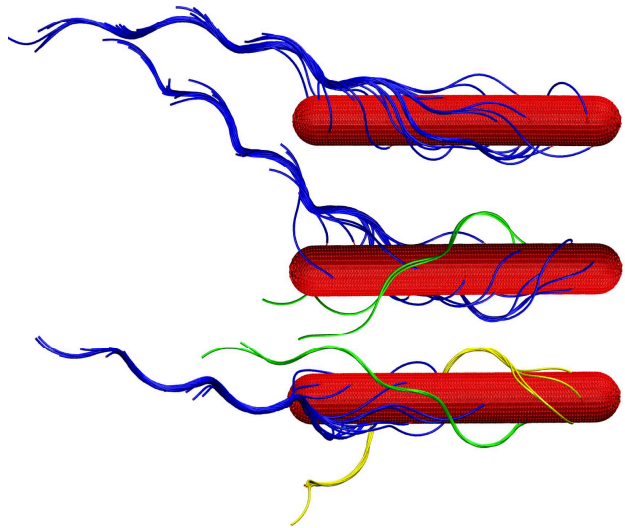}
  \caption{Illustration of cells confined in slits of widths $H/a=20$ (top) and $H/a=120$ (middle, bottom). (top) All flagella are included in a single bundle, (middle, bottom) two and three bundles are formed, respectively. (bottom) The major bundle of $17$ flagella is rather stable for all $H$s as shown in Fig.~\ref{fig:flagella_number}.}
  \label{fig:flagella_snapshot}
\end{figure}

\subsection{Parameters}

We choose $K_{\rm el}^1=K_{\rm el}^2=K_{\rm el}^3=5\times 10^4\, k_BT$, corresponding to a bending stiffness of $2\times 10^{-23}$ N m$^2$ for flagellar filaments within the experimental range of about $10^{-24}-10^{-21}$ N m$^2$. \cite{fuji:72,trac:92,darn:07.1,voge:10,hu:15.1} Moreover, we set
$\vert {\bm T} \vert \le 1000\,k_BT \simeq 4100$ pN nm,\cite{hu:15,hu:15.1} a torque smaller than the stall torque of approximately 4500 pN nm of the flagellar motor of {\em E. coli}. \cite{berr:97}

The cell body is composed of 121 circles with a circle-circle separation of $a/2$. Hence, its  total length is $60a$ and it comprises 3625 mass points. A flagellum contains $N_F =76$ octahedron-like segments with the back-bone bond length $a$, which yields the contour length $76a$. With the pitch angle $30^{\circ}$, the effective length is approximately $66 a$. In total, a cell contains 13125 particles.

The force constants in Eqs.~(\ref{Eq:Vbd}) and (\ref{eq:rep_pot}) are set to $K_{\rm b} = K_{\rm ex}= 10^4 k_BT/a^2$.

The length $a$ of a collision cell, the mass $m$ of a MPC particle, and the thermal energy $k_BT$ define the length, mass, and energy units in our simulations, which yields the unit of time $\tau = a\sqrt{m/k_BT}$.
We choose the collision time step $h = 0.05 \tau$ and average number of fluid particles in a collision cell $\lla N_c\rra=10$, which corresponds to the  fluid viscosity $\eta = 7.15 \sqrt{mk_BT}/a^2$ and the Schmidt number $Sc = 20$.\cite{ripo:05} Newton's equations of motion for the bacterium model are integrated with the time step $ h / 25$ using the velocity-Verlet algorithm. The Reynolds number $Re=l_b m \lla N_c\rra v/ \eta < 0.1$  for the considered body length and velocities (cf. Fig.~\ref{fig:velocity}). Parallel to the walls (cf. Fig.~\ref{fig:sim_angle}), periodic boundary conditions are applied with the box length $160a$, which corresponds to $N = 3.072 \times10^7$ fluid particles for $H/a=120$.

\section{Results---Flagella Bundle}

\begin{figure}[t!]
\centering
  \includegraphics[width=0.45\textwidth]{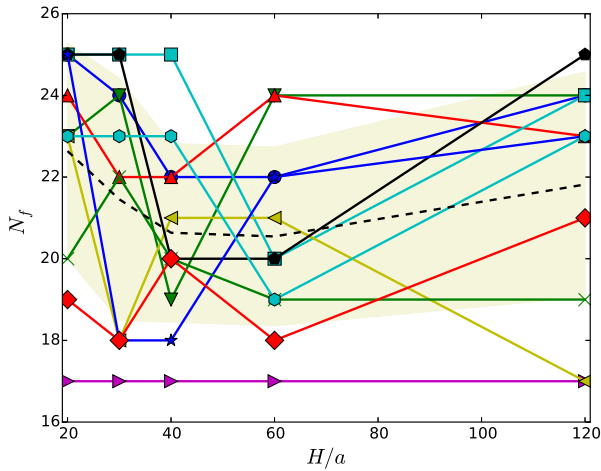}
  \caption{Number of flagella in the major bundle for various wall separations. The dashed line indicates the average over the various realizations labeled by different symbols, and the shaded area the standard deviation.}
  \label{fig:flagella_number}
\end{figure}

\subsection{Bundle Structure}

Hydrodynamic interactions lead to synchronization of the flagella rotation and bundle formation. This aspect of swimming bacteria has been studied in Refs.~\onlinecite{reig:12,reig:13,adhy:15}.
As is evident from the snapshots of Fig.~\ref{fig:flagella_snapshot}, our swarmer cells with, in comparison to swimmers, \cite{hu:15.1} the markedly longer bodies and significantly larger number of flagella also form flagella bundles. Interestingly, typically the majority of flagella  are assembled in a major bundle and only a very few individual flagella or bundles of a few flagella are present. As displayed in Fig.~\ref{fig:flagella_number}, the number of flagella $N_f$ in a bundle depends only weakly on wall separation. With increasing wall separation, only a minor reorganization of the flagella bundle and the number of participating flagella occurs. In the particular case  $N_f=17$, three bundles are formed (cf. Fig.~\ref{fig:flagella_snapshot}), where the number of flagella in the two low-flagella-number bundles fluctuates, but the major bundle is rather stable.

\begin{figure}[t]
\centering
  \includegraphics[width=0.45\textwidth]{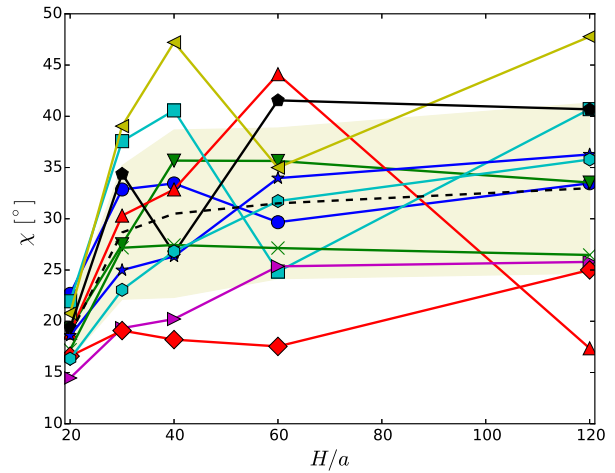}
  \caption{Angle between the cell-body and the bundle major axis. The dashed line indicates the average over the various realizations labeled by symbols. The symbols and colors are the same as in Fig.~\ref{fig:flagella_number}.}
  \label{fig:body_bundel_angle}
\end{figure}

The snapshots of Fig.~\ref{fig:flagella_snapshot} indicate a certain preferred orientation between the cell body and the flagella bundle. Similarly, the images of Ref.~\onlinecite{darn:07.1} suggest such an arrangement of the flagella bundle of swimming {\em E. coli} cells. We characterize this orientation by calculating the angle $\chi$ between the major axis of the body  and the major axis of the moment-of-inertia tensor of the flagella bundle (cf. Fig.~\ref{fig:sim_angle}). Figure~\ref{fig:body_bundel_angle} displays $\chi$ as a function of the wall separation.
Noteworthy, we find a large variation between the various realizations with the same arrangement of flagella on the cell surface, but different initial distributions of velocities, as well as  the various gap widths. We like to emphasize that the variation is not a consequence of insufficient equilibration or sampling. For every individual presented average, the angle $\chi$, as a function of time, moderately fluctuates around a straight line of slope zero and a standard deviation below $\pm 2.5^{\mathrm{o}}$. Only for the narrowest gap, the fluctuations are approximately one degree larger by confinement-induced additional forces. Our studies emphasize that the large variations observed in Fig.~\ref{fig:body_bundel_angle}  are an intrinsic property of self-propelled systems.
Evidently, strong confinement implies a small angle and wall interactions force a more parallel alignment of the body and bundle (cf. Fig.~\ref{fig:flagella_snapshot}). For less confined cells, the angle increases with increasing $H$ and saturates at an $H$-independent value for $H/a>40$. The latter average is almost twice larger than the average of $\chi$ in the narrowest gap.


\subsection{Body and Bundle Rotation}

\begin{figure}[t]
\centering
  \includegraphics[width=0.45\textwidth]{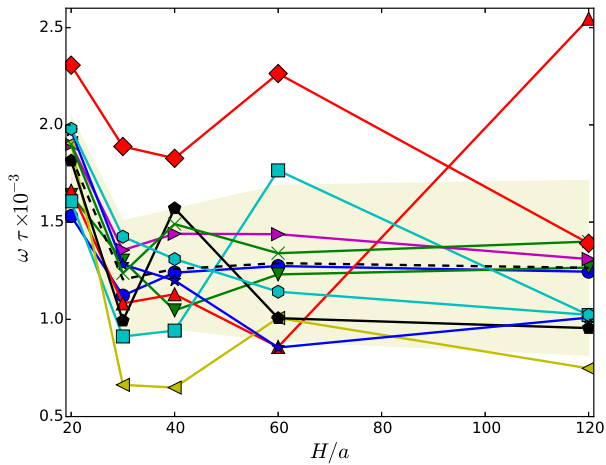}
  \caption{Body rotation frequency projected projected onto the major axis of the cell as a function of the wall separation. The dashed line indicates the average over the various realizations labeled by symbols. The symbols and colors are the same as in Fig.~\ref{fig:flagella_number}.}
  \label{fig:body_rotation_frequency}
\end{figure}

The body-bundle arrangement strongly affects the rotational motion of the cell body. The latter is a consequence of the fact that the cell is torque free, i.e., the induced rotation of the flagella implies the well-known counter rotation of the cell body. \cite{darn:07.1} We determine the rotation frequency $\bm \omega_b$ of the cell body via the relation $\bm L_b = \bm {\Theta}_b \bm \omega_b $, where the angular momentum $\bm L_b$ of the body and its moment of inertia tensor  $\bm {\Theta}_b$ are given by
\begin{align}
\bm L_b = & \sum_i M \Delta \bm r_i \times \Delta \bm v_i  , \\
{\Theta_b}_{\alpha \beta}  = & \sum_i M \left( \delta_{\alpha \beta}\Delta \bm r_i^2  - \Delta {r_i}_\alpha \Delta {r_i}_\beta  \right) .
\end{align}

The $\Delta \bm r_i$ and $\Delta \bm v_i$ are the positions and velocities of the particles comprising the body  with respect to the center-of-mass position and velocity, respectively, of the body. Figure~\ref{fig:body_rotation_frequency} displays the body rotational frequency projected onto the major axis of the cell, i.e., $\omega= \bm \omega_b \cdot \bm e$, where $\bm e$ is a unit vector along the major axis of the inertia tensor of the whole cell (body plus flagella). The rotation frequency is virtually constant for $H/a \gtrsim 30$. Only the strongest confined cells exhibit a by a factor of two larger $\omega$. For the wider gaps, the distribution of $\omega$ is rather broad; comparable with the mean value itself.

As expected, the $\omega$ values are strongly linked with the body-bundle angle $\chi$---an increase of $\chi$ causes a decrease of $\omega$. In a straight configuration of the body and the bundle, both (counter) rotate essentially around the major axis of the body. An increase of the angle $\chi$ implies an additional rotation, a wobbling motion, of the whole cell around an oblique axis none aligned with $\bm e$, hence, $\omega$ is smaller.

\section{Results---Swarmer Distribution Between Walls}

\begin{figure}[h!]
\centering
   \includegraphics[width=0.9\columnwidth]{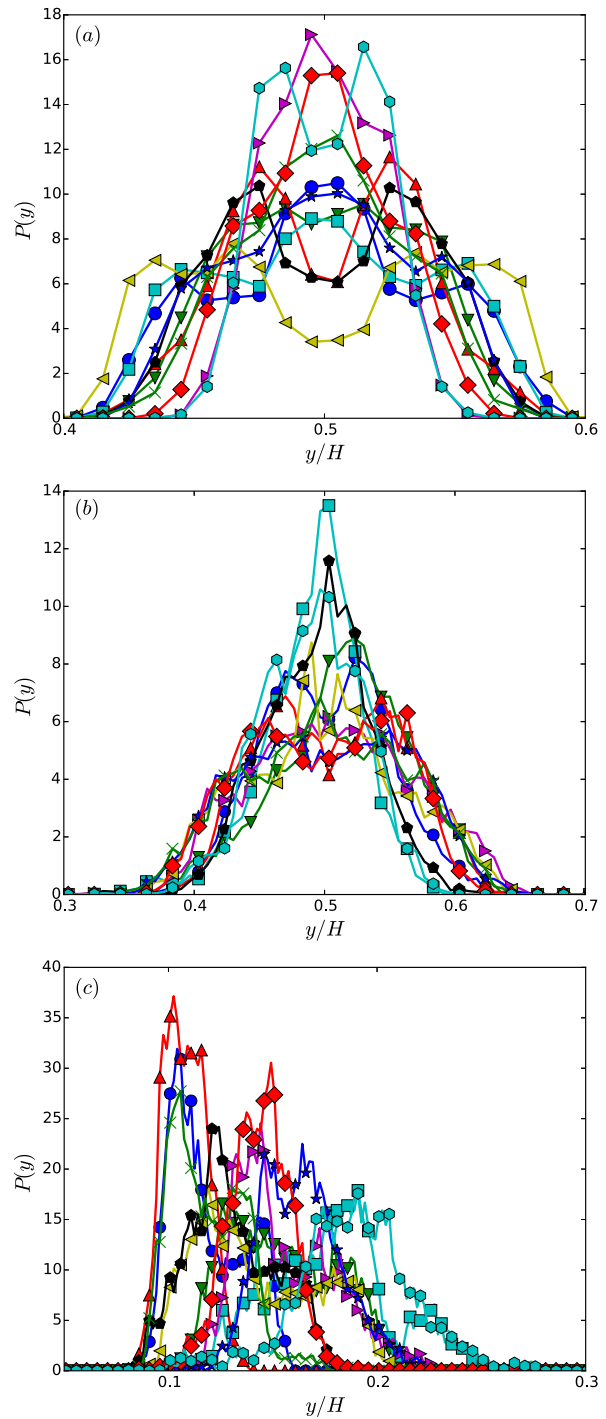} \\
  \caption{Distribution functions of the center-of-mass position of cell bodies for various realizations and the wall separations (a) $H/a = 20$ or $H/l_b = 1/3$, (b) $H/a=30$ or $H/l_b = 1/2$, and (c) $H/a=120$ or $H/l_b = 2$.  The symbols and colors are the same as in Fig.~\ref{fig:flagella_number}.}
  \label{fig:distributions}
\end{figure}

Figure~\ref{fig:distributions} shows distribution functions $P(y)$ of the center-of-mass position of the cell body between the two walls. They are normalized such that $\int_0^H P(y) dy =1$. In the narrowest gap $H/a=20$, the distributions are symmetric with respect to the gap center, however, $P(y)$ varies substantially between the various realizations. Note that the length of the cell body is $l_b/a = 60$ and its diameter is $d_b/a=9$.  Cells with small angles $\chi$ exhibit a major peak in the center between the walls. With increasing angles, two off-center peaks appear, and for large $\chi$ off-center peaks close to the walls develop together with a central peak. The realization labeled by the yellow $\blacktriangleleft$ symbol shows an even more distinct distribution, although the differences in the flagella arrangement is a priori not evident.
At $H/a=30$, the cells are mainly concentrated in the gap center (Fig.~\ref{fig:distributions}(b)). Still, there are realizations, which prefer walls, giving rise to off-center peaks.
In the infinite time limit, symmetric distributions are expected and will occur. Since we are able to only average over a limited time, the present asymmetry in the distributions reflects particular long-lived structures.

The swarmer cells in a gap of width $H/a=120$, i.e., $H/l_b=2$, reveal a preference to reside near a wall for a long time. The cells are preferentially located within a layer of half a body length adjacent to the wall, or approximately three body diameters. Although, the cells are initially close to the wall, the simulation time is long enough such that the cells could diffuse a larger distance. Considering the time dependence of the cell height above the wall, only the cell furthest apart from the wall (cyan hexagons) exhibits a trend to move away from the wall. The other realizations are rather stable over the considered time range, again, indicating long-lived structures.
The actual alignment between body and bundle seems to be of minor importance for the attachment close to the wall.  By calculating the angle between the cell-orientation vector and the surface normal, we find  that the cells are preferentially oriented  toward the nearby wall during the simulations, with average angles in the range $-2^{\circ}$ to $-10^{\circ}$. In the stationary state, the distribution functions will be symmetric with respect to the center of the gap, with equal probability to find the cells next to either of the walls. To reach this state is far beyond current computer simulations capabilities, when rather detailed cell models are used as in our study. Nevertheless, our simulations shed light onto temporarily stable and long-lived behaviors.

The preference to stay close to a wall is attributed to hydrodynamic interactions. \cite{frym:95,elge:15,berk:08,laug:12,hu:15,shum:15,dres:11,swie:13,lint:16} A reduction of the wall separation leads to an overlap of the effective attraction of the two walls, which implies a preference of the cells to stay in the center of the gap. This has already been observed in theoretical calculations \cite{shum:15} and experiments. \cite{swie:13} The theoretical considerations in Ref.~\onlinecite{shum:15} of monotrichous cells with well-aligned cell body and flagellum ($\chi=0$) show a crossover from single-wall behavior of cells in gaps of width $H/l_b \approx 4$ to a preference in the gap center for $H/l_b < 1$. The theory predicts stable fixed points in front a wall for large wall separations. \cite{shum:15} These fix points become unstable at small wall separations and an initially unstable fixed point in the gap center becomes stable. Similarly, the experiments of Ref.~\onlinecite{swie:13} indicate a stable position in the gap center for $H/l_b<1$. Our simulation results are qualitatively consistent with these findings. However, we predict a certain influence of the flagella bundle arrangement on the preferred location in the gap as long as the gap width is comparable to the cell diameter. Aside from hydrodynamic interactions, steric interactions between a cell and the walls matter. This is particularly pronounced for very narrow gaps as shown in Fig.~\ref{fig:distributions}(a).

\section{Results---Swarmer Dynamics}

\begin{figure}[t]
\centering
  \includegraphics[width=0.45\textwidth]{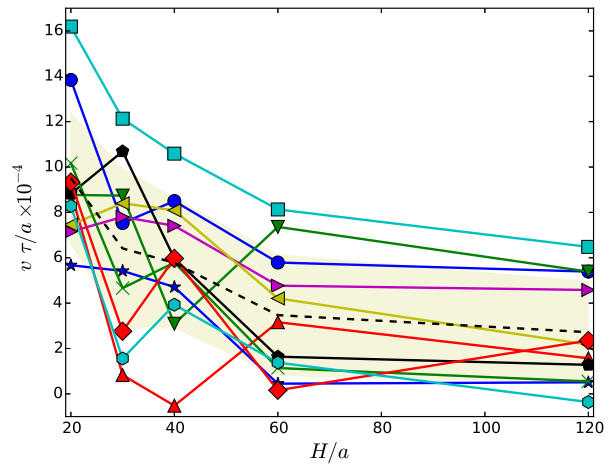}
  \caption{Migration velocity of swarmer cells confined between two walls for various realizations. The dashed line indicates the average over the various realizations labeled by symbols. The symbols and colors are the same as in Fig.~\ref{fig:flagella_number}.}
  \label{fig:velocity}
\end{figure}

To characterize the motility of the cells, we calculate their migration velocity $v$ and the their mean square displacement. The instantaneous  velocity  $v(t)$ is defined as follows. We calculate the displacement $\Delta \bm r_b(t) = \bm r_b(t+\Delta t) - \bm r_b(t)$ of the center of mass of the body for the lack time $\Delta t$. Dividing by $\Delta t$ yields the velocity $\bm v_b(t)$. We then project this velocity onto the major axis of the inertia tensor of the whole cell, i.e., $v=\bm v_b \cdot \bm e$. Results of the average velocity $v$ for the various gap widths are displayed in Fig.~\ref{fig:velocity}. Obviously, the velocity decreases with increasing wall separation until a virtually $H$ independent value is assumed for $H/a\gtrsim 60$. Thereby, cells in wider gaps are on average by a factor of three slower than strongly confined cells. This can partially be attributed to the body-bundle orientation, especially for  $H/a=20$, but for wider gaps, the angle $\chi$ is nearly constant whereas $v$ decreases further. Hence, the change in velocity seems to be related to wall-cell hydrodynamic effects.

\begin{figure}[h]
\centering
  \includegraphics[width=0.45\textwidth]{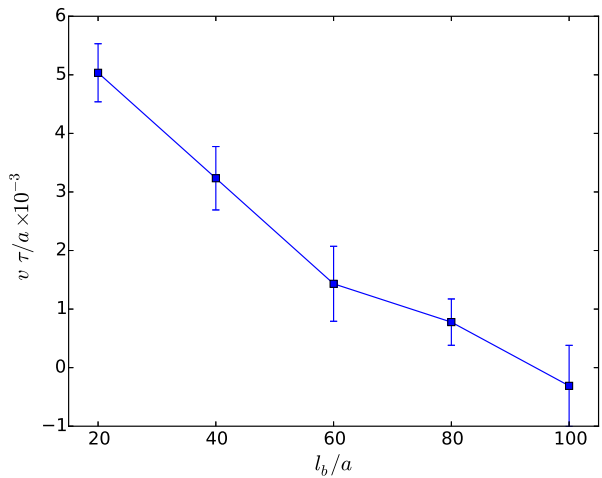}
  \caption{Migration velocities of swarmer cells confined between two walls of separation $H/a=20$  as function of their body length.  The number of flagella increases linearly with $l_b$ starting from $N_f=4$ for $l_b/a=20$.}
  \label{fig:velocity_length}
\end{figure}

\begin{figure}[h]
\centering
  \includegraphics[width=0.45\textwidth]{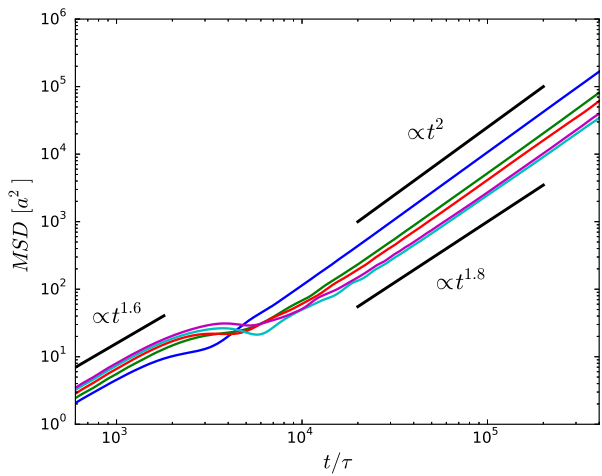}
  \caption{Mean square displacements of the center of mass of the body for cells confined in gaps of widths $H/a=20$ (blue), $H/a=30$ (green), $H/a=40$ (red), $H/a=60$ (cyan), and $H/a=120$ (purple).  }
  \label{fig:mean_square_displacement}
\end{figure}

The migration velocity depends not only on the extend of confinement, but also on the body size. This is illustrated in Fig.~\ref{fig:velocity_length}. Note, in Fig.~\ref{fig:velocity_length}, the flagella density is smaller than in the other studies  presented in the article. The  dependence of the migration velocity on the flagella number remains to be studied. Evidently, the velocity decreases significantly with increasing body-length and assumes rather small values for long cells. The error bar indicates that even backwards-swimming realization appear. In general, the decrease of the velocity $v$ with increasing body length in narrow slits is similar to the behavior of such cells in bulk fluids. The initial linear decrease  of the velocity $v$ is consistent with theoretical expectations based on resistive force theory.\cite{elge:15,laug:09,rode:13,chat:06,maga:95,purc:97} Here, the friction of the cell body dominates the resistance in the migration and, hence, $v \sim 1/l_b$.  However, this is in contrast to experimental results, where the mean speed of a planktonic {\em E. coli} cell is comparable to that of an elongated planktonic cell. \cite{swie:13} Our simulations predict a factor of three larger migration velocity of planktonic cells. The reason of this large discrepancy between experiment and simulation is not evident and needs to be further studied. However, obviously  a stronger torque of the molecular motors exerted on a flagellum with a respective higher flagellum rotation frequency
would lead to a faster mean speed. It needs to be clarified whether flagella in swarmer or elongated planktonic cells exhibit such an increased rotation frequency.

\begin{figure}[t]
\centering
  \includegraphics*[width=0.45\textwidth]{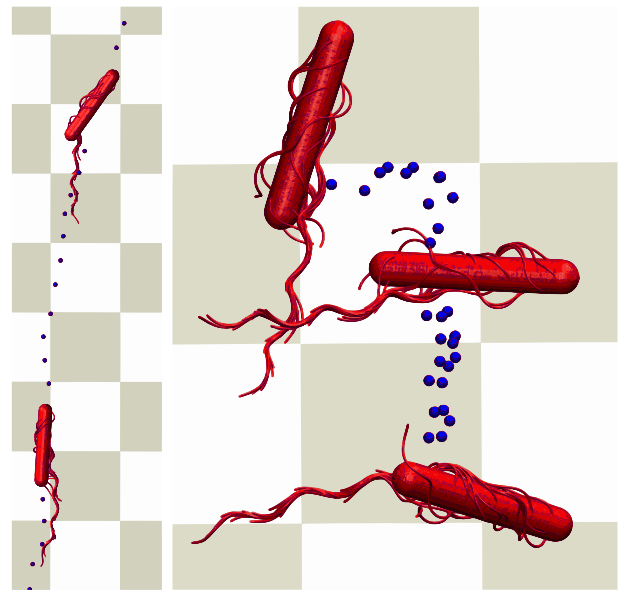}
  \caption{Illustration of migration patterns. The left cell ($H/a=20$) migrates rather straight (bottom to top), whereas the right cell ($H/a=120$) rolls over the wall (top to bottom). The dots indicate subsequent positions in time of the center of mass of the cell. Simulation
animations are shown as movie S1 and S2 in ESI.}
  \label{fig:cell_migration}
\end{figure}

\begin{figure}[t]
\centering
\includegraphics[width=0.45\textwidth]{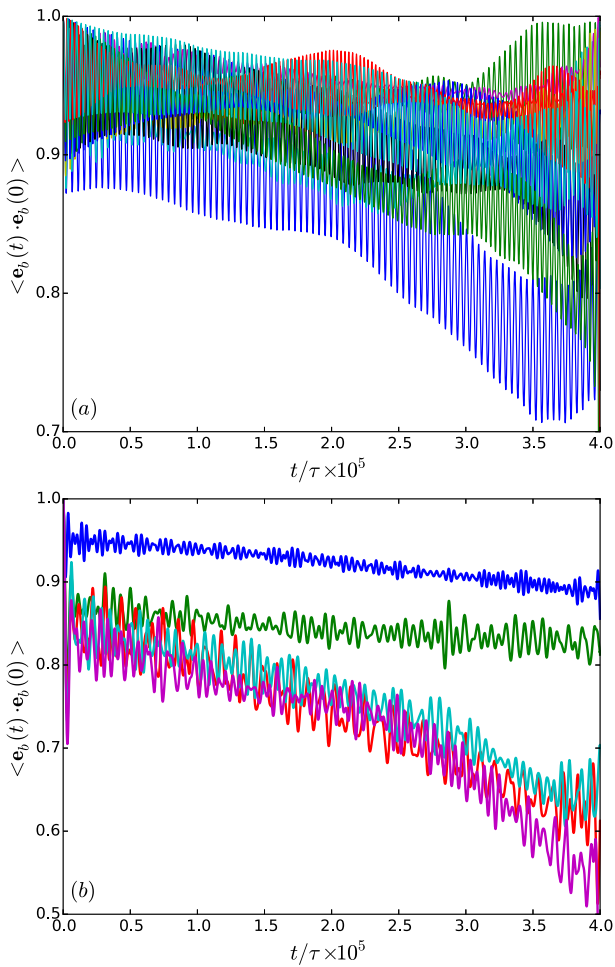}
  \caption{(a) Body-orientation correlation functions of cells confined in a film of width $H/a=20$ for various realizations. (b) Body-orientation correlation functions averaged over the various realizations for  $H/a=20$ (blue), $H/a=30$ (green), $H/a=40$ (red), $H/a=60$ (cyan), and $H/a=120$ (purple).  }
  \label{fig:orientation_corr}
\end{figure}

The mean square displacement (MSD) of the center of mass of the cell body is presented in Fig.~\ref{fig:mean_square_displacement}. For a given wall separation, two regimes can be identified. At short times $t/\tau \lesssim 10^{3}$, the MSD increases in a superdiffusive manner as $t^{1.6}$ mainly due to the inertia of the cell, i.e., activity little affects the dynamic in this time regime. Thereby, there is little variation between the various gap widths. When the cell displacement, the square root of the MSD,  reaches approximately a cell diameter at $t/\tau \approx 2 \times 10^3$,  the MSD assumes a plateau-like value. This behavior is attributed to the {\em wobbling} motion of the cell body. While for the very narrow gap, wobbling is suppressed by  steric interactions with the surfaces, it is well pronounced for wider gaps. The wobbling motion is illustrated in the simulation animations S1 and S2 of the ESI.

At times longer than about $t/\tau =  10^4$, another superdiffusive regime is assumed. In particular, the cells confined in the narrowest gap exhibit a ballistic motion. This is related to rather straight trajectories of the individual cells (cf. Figure~\ref{fig:cell_migration}). For the wider gaps, the MSD increases somewhat slower, but it is close to ballistic motion. The smaller exponent points toward a modified migration behavior of the cells, possibly by curved trajectories or other types of migration behavior. In general, however, there are seemingly only minor qualitative differences between the various slit widths over the considered range of displacements on the order of $5 l_b$. In the asymptotic limit $t\to \infty$, we expect that the cells exhibit a diffusive motion. However, to reach this regime requires much longer simulations.

In order to characterize the orientational stability of trajectories, we determine the body orientational correlation function
\begin{align} \label{eq:body_correlation}
C_b(t) =\lla \bm e_b(t) \cdot \bm e_b(0) \rra ,
\end{align}
with $\bm e_b$ the unit vector along the major body axis.
Results for the narrow gap $H/a=20$ are displayed in Fig.~\ref{fig:orientation_corr}(a).  In general, the orientations are rather persistent and only some realizations exhibit a moderate decay of $C_b(t)$.  The various realizations exhibit fast oscillations, which correspond to the frequencies presented in Fig.~\ref{fig:body_rotation_frequency}. Moreover, certain slower oscillations are superimposed, indicating a rich overall dynamics of a cell due to the angle between the body and the bundle. Correlation functions averaged over the various realization are presented in Fig.~\ref{fig:orientation_corr}(b). In the narrower gaps with $H/a=20$ and $30$, the cells swim in a rather straight manner and the correlation functions decay slowly. This is illustrated in the snapshots of Fig.~\ref{fig:cell_migration}. For slit widths $H/a \gtrsim 40$, $C_b(t)$ decays rather similarly for the considered cases. Thereby, the correlation function decays significantly faster compared to narrower gaps. This is related to the single-wall behavior of the cells for the wider slits.  Here, the dynamical pattern is quite heterogeneous. Some of the cells start to move along circles, similar to planktonic cells, \cite{hu:15,laug:06,dile:11,leme:13}
where we are only able to see part of the circular path during the considered simulation time.
Other cells rather roll over the surface as illustrated in Fig.~\ref{fig:cell_migration}. This is partially caused by the, in case of our simulations, lower migration speed of swarmer cells compared with planktonic cells. More importantly, the flow profile of swarmer cells is more complex in comparison with that of swimming bacteria, \cite{hu:15.1} since the body is in part covered with flagella bundle(s). The interplay between the rotating bundle (behind the cell body) and the counterrotating flagella-covered cell body can lead to cell-rolling in a curved manner (cf. Fig.~\ref{fig:cell_migration}).
Hence, in combination with wobbling, we observe a distinctively different surface behavior of the considered swarmer cells due to the rotating flagella bundle and the counterrotating cell body.  Rolling over a surfaces has been observed experimentally for artificial bacteria flagella.\cite{peye:10}

In general, the correlation functions decay in an non-exponential manner. This is certainly not surprising, since the rotational diffusion coefficient is low and we did not cover the time scale corresponding to the inverse of this diffusion coefficient. Nevertheless, the results of Fig.~\ref{fig:orientation_corr}(b) reveal a qualitative difference in the migration behavior in rather narrow or wider gaps.

\section{Summary and Conclusions}

We have proposed a model for an {\em E. coli}-type swarmer cell and have investigated its properties in confinement between two parallel walls. In general, we find a significant heterogeneity in the appearing structures and the dynamics. For very narrow gaps, where $H/l_b \lesssim 1/3$, confinement strongly affects the bundle arrangement and the dynamical properties of cells such as the cell rotation frequency and the migration velocity. For gap widths $H/l_b \gtrsim 2/3$, $H$ independent values are assumed. The distribution of cells sensitively depends on the wall separation. In very narrow gaps ($H/l_b=1/3$), the bundle arrangement matters and geometrical restrictions are essential. With increasing wall separation ($H/l_b=1/2$), both walls become equally important and the cells migrate in the center between the walls, whereas for large $H/l_b \gtrsim 1$ cells stay close to one of the walls over the simulation time. Considering the migration patters, we find straight paths for narrow gaps, but also rolling over a surface of cells for wide gaps. This is reflected in the cells' center-of-mass mean square displacement.

Our calculations show a decrease of the cell migration velocity  with increasing body length. This result is in agreement with theoretical expectations within the resistive force theory.\cite{elge:15,laug:09,rode:13,chat:06,maga:95,purc:97} However, Ref.~\onlinecite{swie:13} states that elongated {\em E. coli} planktonic cells migrate as fast as planktonic {\em E. coli} cells and that {\em E.coli} swarmer cells migrate even $60\%$ faster. This clearly contradicts our simulation results and theoretical expectations. It remains to be clarified, why swarmer cells migrate so efficiently. Hypothetically, an increased torque in the elongated cells would give rise to faster migration.

Interestingly, essentially a single bundle is formed including in average approximately $80\%$ of the flagella, which gives rise to a migration behavior rather similar to swimming planktonic cells. This is related to the length ratio between the cell body and the flagella; in our case the ratio is approximately unity. Such cells exhibit collective swarming behavior as seen in various experiments.\cite{cope:09,darn:10,kear:10,part:13} However, significantly longer swarmer cells, longer than the length of a flagellum, form multiple bundles and are expected to show a distinctively different swarming behavior. Specifically, inter-cell bundles might be formed as suggested in Refs.~\onlinecite{stah:83,jone:04}. We are currently working on an extension of our studies to such longer cells.

Finally, we would like to stress once more the importance of hydrodynamic interactions for bundle formation.\cite{reig:12,reig:13} Tests confirm that bundles are also occasionally formed without hydrodynamic interactions\cite{adhy:15} due to flagella rotation and the associated counter-rotation of the cell body, but that strongly depends on the initial arrangement of flagella. However, in presence of hydrodynamic interactions, bundles are always formed.

\section*{Acknowledgments}
We thank G. Gompper for helpful discussions.
Financial support by the Deutsche Forschungsgemeinschaft (DFG) within the priority program SPP 1726 ``Microswimmers -- from Single Particle Motion to Collective Behaviour'' is gratefully acknowledged.

\footnotesize{
\bibliographystyle{rsc} 
\providecommand*{\mcitethebibliography}{\thebibliography}
\csname @ifundefined\endcsname{endmcitethebibliography}
{\let\endmcitethebibliography\endthebibliography}{}

}

\end{document}